\documentclass{sigchi}

\CopyrightYear{2018} 
\setcopyright{acmlicensed} 
\conferenceinfo{CHI 2018,}{April 21--26, 2018, Montreal, QC, Canada}
\isbn{978-1-4503-5620-6/18/04}
\acmPrice{\$15.00}
\doi{https://doi.org/10.1145/3173574.3174130}

\usepackage{balance}       
\usepackage{graphics}      
\usepackage[T1]{fontenc}   
\usepackage{txfonts}
\usepackage{mathptmx}
\usepackage[pdflang={en-US},pdftex]{hyperref}
\usepackage{color}
\usepackage{booktabs}
\usepackage{textcomp}

\usepackage{microtype}        
\usepackage{ccicons}          

\usepackage{todonotes}

\usepackage{amsmath}
\usepackage{caption}
\usepackage{subcaption}
\usepackage{multirow}
\usepackage{makecell}
\usepackage{arabtex}
\usepackage[american]{babel}
\usepackage{csquotes}
\usepackage{array} 
\usepackage{siunitx}
\newcolumntype{L}{@{}l@{}} 
\usepackage[font={small}]{caption}
\usepackage{utf8}

\def\plaintitle{Measuring, Understanding, and Classifying News Media Sympathy on Twitter after Crisis Events}

\def\emptyauthor{}
\def\plainkeywords{Twitter; sympathy; sentiment analysis; news media bias; crisis informatics; cross-cultural; crowdsourcing; NLP}

\makeatletter
\def\url@leostyle{%
  \@ifundefined{selectfont}{
    \def\UrlFont{\sf}
  }{
    \def\UrlFont{\small\bf\ttfamily}
  }}
\makeatother
\def\pprw{8.5in}
\def\pprh{11in}

\definecolor{linkColor}{RGB}{6,125,233}
\hypersetup{%
  pdftitle={\plaintitle},
  pdfauthor={\emptyauthor},
  pdfkeywords={\plainkeywords},
  pdfdisplaydoctitle=true, 
  bookmarksnumbered,
  pdfstartview={FitH},
  colorlinks,
  citecolor=black,
  filecolor=black,
  linkcolor=black,
  urlcolor=linkColor,
  breaklinks=true,
  hypertexnames=false
}

\begin{document}

\title{\plaintitle}

%

\numberofauthors{1}
\author{
  \alignauthor Abdallah El Ali$~^{1,2}$, Tim C Stratmann$~^{2}$, Souneil Park$~^{3}$, \\ Johannes Sch{\"o}ning$~^{4}$, Wilko Heuten$~^{5}$, Susanne CJ Boll$~^{2}$\\
    \affaddr{$^{1}$~ Centrum Wiskunde \& Informatica (The Netherlands), $^{2}$~University of Oldenburg (Germany), $^{3}$~Telefonica Research (Spain), $^{4}$~University of Bremen (Germany), $^{5}$~OFFIS - Institute for IT (Germany)}\\
    \email{a.elali@cwi.nl, tim.claudius.stratmann@uol.de, souneil.park@telefonica.com, schoening@uni-bremen.de, heuten@offis.de, susanne.boll@uol.de}   
}

\maketitle

\begin{abstract}
This paper investigates bias in coverage between Western and Arab media on Twitter after the November 2015 Beirut and Paris terror attacks. Using two Twitter datasets covering each attack, we investigate how Western and Arab media differed in coverage bias, sympathy bias, and resulting information propagation. We crowdsourced sympathy and sentiment labels for 2,390 tweets across four languages (English, Arabic, French, German), built a regression model to characterize sympathy, and thereafter trained a deep convolutional neural network to predict sympathy. Key findings show: (a) both events were disproportionately covered (b) Western media exhibited less sympathy, where each media coverage was more sympathetic towards the country affected in their respective region (c) Sympathy predictions supported ground truth analysis that Western media was less sympathetic than Arab media (d) Sympathetic tweets do not spread any further. We discuss our results in light of global news flow, Twitter affordances, and public perception impact.

\end{abstract}

\category{H.5.3.}{Group and Organization Interfaces}{Web-based interaction}{}{}

\vspace{-0.5pc}
\keywords{\plainkeywords}

\section{Introduction}

On 12 November, 2015, the city of Beirut (Lebanon) witnessed two bombings\footnote{https://en.wikipedia.org/wiki/2015\_Beirut\_bombings; last retrieved: 24.12.2017} at approximately 18:00 Eastern European Time (EET) (UTC+02:00), coordinated by two suicide bombers that detonated explosives in Bourj el-Barajneh, a southern suburb of Beirut. This suburb is largely inhabited by Shia Muslims, and reports estimate the number of deaths to be anywhere between 37 to 43, with over 200 injured. These bombings constituted the worst attack in Beirut since the end of the 1990 Lebanese Civil War. Shortly after the attacks, the Islamic State (IS)\footnote{In Arabic also known as Da'ish.} claimed responsibility for the attacks.

A day later, on 13 November, 2015 beginning at 21:20 Central European Time (CET) / UTC+01:00, three suicide bombers carried out a series of coordinated terrorist attacks in Paris\footnote{https://en.wikipedia.org/wiki/November\_2015\_Paris\_attacks; last retrieved: 24.12.2017} on its northern suburb, Saint-Denis. They struck near the Stade de France in Saint-Denis, followed by suicide bombings and mass shootings at caf\'{e}s, restaurants and a music venue in central Paris. The attacks resulted in the deaths of 130 people, and injury of 368 people. These attacks were purported to be the deadliest on France since World War II. Shortly after the attacks, IS also claimed responsibility for the attacks.

The days after the November attacks in Paris, people took social media by storm in response to the events. From outcries of sympathy and solidarity with Paris, to outcries against or support for Islam \cite{Magdy2016}, to proclamations that mainstream coverage of the Beirut attacks have been sparse and uncaring. In such press reports, we find allegations suggesting reports about attacks, bombings and other crisis events that Western media do not sympathize with attacks in the Arab world as much as they do for attacks in the Western world. Even though many studies have already shown that news providers are inherently biased (such as in what to cover) \cite{Baron2006,Herman1988} where attention asymmetries exist \cite{Abbar2015}, the question arises why differential coverage across crisis events may occur in social media content.

According to Kwak and An \cite{Kwak2014}, how a piece of news is reported (specifically how sympathetic it is to those affected) has the capacity to evoke compassion, which can facilitate monetary support  and launch collective public action. This has far reaching implications given the media's power (even on microblog platforms like Twitter) in shaping public discourse and perception of global events \cite{Vincent1997}, which raises the following questions: To what extent cultural, religious, or geo-political factors account for news media bias in coverage of global terror attacks? And what makes a country newsworthy of sympathy under such events? We believe this has implications for how news organizations can better manage public opinion using social media sites during the immediate and long-term aftermath of a political and/or religious crisis event (e.g., terror attack). Importantly, exposure to biases has been shown to have the capability to foster intolerance and create ideological segregation in major political and social issues \cite{Glynn2004}, and this may be amplified across Western and Arab cultures. Considering the time-critical nature of Twitter news reporting, it becomes key to develop approaches for understanding sympathy, and as we show later, our work contributes a first step towards such an approach for viewing Twitter as a journalism tool that can help in detecting sympathy.

During November 2015, examples of online news headlines included: \textit{``Beirut, Also the Site of Deadly Attacks, Feels Forgotten"}\footnote{http://www.nytimes.com/2015/11/16/world/middleeast/beirut-lebanon-attacks-paris.html; last retrieved: 24.12.2017} by The New York Times, \textit{``Paris, Beirut, and the Language Used to Describe Terrorism"}\footnote{http://www.theatlantic.com/international/archive/2015/11/paris-beirut-media-coverage/416457/; last retrieved: 24.12.2017} by The Atlantic, or \textit{``Did the media ignore the Beirut bombings? Or did readers?"}\footnote{http://www.vox.com/2015/11/16/9744640/paris-beirut-media; last retrieved: 24.12.2017} by Vox. While coverage of the attacks may have been disproportionate, discussion as to why this was so was polarized. On one end, public and journalistic response attributed blame towards the media and its volume and style of coverage (cf., The Atlantic), and on the other end (cf., Vox) there were claims that the media did its part in adequate news reporting, but since Western readers did not care, coverage was drastically reduced. Given such statements, this paper tries to quantify the differences across this observed differential news coverage.

We believe these events provide an interesting use-case to study, because: (a) it allows us to examine in detail whether Western and Arab media exhibit differences in reporting global crisis events, and if so, how such differences manifest (b) the unique context of our study, marked by the rare successive occurrences of two crisis events at places within the Western and Arab world, allows exploring the relative differences specific to those two worlds (c) it allows us to examine such biases using publicly available social media data (in our case, Twitter), and in turn how this can aid journalistic social media practice in ensuring transparency and quality in produced content. 

In this paper, we adopt research on media bias to study news on Twitter (cf., \cite{Golbeck2011,Wei2013}). Our focus here is not on whether media biases exist within Twitter as a social media platform specifically, but rather we use Twitter as a journalism tool to study time-critical news reporting and provide first steps to develop an approach for classifying sympathetic tweets. Here, the affordances of Twitter as a platform can be redesigned to aid news reporters by automatically detecting (sympathy) biases. Twitter data was deemed suitable because: (a) Twitter has become a popular channel for news reporting by mainstream media \cite{Kim2015,Moon2014} (b) the large quantity of tweets allows us to gain insight into the temporal aspect of news coverage at a finer grained level than with news articles (c) the uncensored nature of tweets that are collected with the Twitter Streaming API (d) the costs involved in labeling tweets for sympathy are lower than lengthier pieces of news coverage.

\section{Research Questions \& Contributions}

Given the foregoing motivation, we aim to better understand media coverage differences on Twitter through computationally capturing news sympathy during such unexpected, human-induced crisis events. Bias can be viewed as a partial perspective on facts \cite{Shoemaker1996}, which can be further broken down into three aspects \cite{Alessio2000}: selection bias (gatekeeping), or which stories are selected; coverage bias, or how much attention is given to a story; and statement bias, or how a story is reported. Here, we used these concepts and zoomed in on the Beirut and Paris attacks, to examine in detail whether their coverage on Twitter differed. Here, we focus on coverage bias, statement bias (specifically on characterizing news sympathy), and whether sympathetic messages propagate further on Twitter. Since we are only concerned with the differences in reporting of these two events, we do not consider selection bias. The underlying assumption we make here is that it is important to minimize such bias, even if only on social networks like Twitter. We posit the following questions: 

\begin{itemize}
\item \textbf{RQ1 - Coverage bias:} Was there a difference (in terms of normalized tweet volume) between Western and Arab media coverage of the Beirut and Paris attacks?
\item \textbf{RQ2 - News media sympathy bias:} Was there a difference between Western and Arab media in how sympathetic the tweets were towards affected individuals across each event?
\item \textbf{RQ3 - Information propagation:} Do sympathetic and positive sentiment tweets propagate further throughout the Twitter network (i.e., receive more retweets)?
\end{itemize}

For coverage bias, we hypothesized that the Beirut attacks would receive less coverage from Western media, but not from Arab news media accounts on Twitter, with the inverse for coverage of Paris. This is in line with the news flow theory \cite{Segev2015}, which states that the prominence of a foreign country in the news is attributed to three groups of variables: (a) national traits (e.g. size and power of the foreign country), (b) relatedness (e.g., proximity to a foreign country in terms of geography, demography, etc.) and (c) events (e.g., wars, disasters, protests) \cite{Segev2015,Wu2000}. In this case, Paris is both geographically and culturally closer to Western countries, and given the timeline of both attacks, Paris would likely attract more coverage. 

Given news statements (e.g., NYTimes's article\footnote{http://www.nytimes.com/2015/11/16/world/middleeast/beirut-lebanon-attacks-paris.html; last retrieved: 24.12.2017}) on differential coverage and Diakopoulos's \shortcite{Diakopoulos2014} work on Twitter newsworthiness, we expected that tweets from Western media covering the Beirut attacks would overall exhibit less sympathy than coverage of the Paris attacks, while the inverse is true for Arab Twitter news media, which would exhibit more sympathy towards the Beirut attacks. In this paper, our objective is to explore these questions using a combination of NLP techniques and crowdsourcing on Twitter datasets. Finally, we look at information diffusion \cite{Starbird2012}, where previous research has found that negative news sentiment on Twitter enhances virality \cite{Hansen2011}, while other work found that emotionality and positivity in online news content increases virality \cite{Berger2012}. Despite mixed findings, we expected that tweets which are sympathetic (as well as ones with positive sentiment) to be more likely to spread throughout the Twitter network, by resulting in more retweets.

In this paper, we make the following contributions to human-computer interacton and social computing research:

\begin{enumerate}

\item We provide a deeper, data-driven understanding of how the Western and Arab news coverage of the Paris and Beirut attacks differed with respect to coverage bias, news sympathy, and information propagation.

\item We provide a public annotated multi-language (English, Arabic, French, German) dataset that can be used to train learning algorithms to predict news sympathy on Twitter during terror events (see Supplementary Material A\footnote{\url{https://github.com/abdoelali/CrisisNewsSympathy}}).

\end{enumerate}

\section{Related Work}

\subsection{Media Bias in Communication and Social Media}

Trumper et al. \cite{Trumper2013} examined biases in online news sources and social media communities around them, and by analyzing 80 international news sources during a two-week period, they showed that biases are subtle but observable, and follow geographical boundaries more closely than political ones. Sert et al. \cite{Sert2016} proposed to leverage user comments along with the content of the online news articles to automatically identify the latent aspects of a given news topic, to be used as a first step in detecting the news resources that are biased towards certain subsets of these latent aspects. Park et al. \cite{Park2009} took a different approach towards media bias with NewsCube, where they  automatically provide readers with multiple classified viewpoints on a news event of interest. Dallmann et al. \cite{Dallmann2015} investigated a dataset covering all political and economical news from four leading online German newspapers over four years, and showed that statistically significant differences in the reporting about specific parties can be detected between the analyzed online newspapers.

More generally, social media content biases have been observed across datasets. In a study of Twitter, Flickr, and Swarm, it was shown that volunteered geographic information is ``local" (geotags correspond closely with the home locations of its contributors) in only about 75\% of cases \cite{Johnson2016}, and compounded by how localness is defined \cite{Kariryaa2018}. This effect influences the design of geolocation inference algorithms, which have been shown to exhibit significantly worse performance for underrepresented populations (i.e. rural users), even when overcorrecting for populations biases \cite{Johnson2017}.

\subsection{Twitter for Crisis and Controversy Understanding}

Twitter has shown to be a rich resource to study media bias and controversy, especially after major events, whether political, religious, or natural (e.g., \cite{An2016,Holthoefer2015,Shamma2009}). Morgan et al. \cite{Morgan2013} found that Twitter users share news in similar ways regardless of outlet or perceived outlet ideology, and that as users share more news content, they tend to quickly include outlets with opposing views. Younes et al. \cite{Younus2012} looked at how traditional media outlets and social media differ in coverage of an event, and focused on coverage patterns of NYTimes articles and tweets during the Egyptian uprising in 2011. To discover such patterns, they proposed a simple media bias measurement model for day-to-day news items built on top of topic models. They found that traditional news sources have a wider disparity in the ranks and hence a strong presence of media bias. 

Wei et al. \cite{Wei2013} proposed an empirical measure to quantify mainstream media bias based on sentiment analysis and showed that it correlates better with the actual political bias in the UK media than pure quantitative measures based on media coverage of various political parties. They then studied media behavior on Twitter during the 2010 UK General Election, and showed that while most information flow originated from the media, they seem to lose their dominant position in shaping public opinion during this general election. Olteanu et al. \cite{Olteanu2015} investigated several crises using public Twitter data -- including natural hazards and human-induced disasters -- in a systematic manner and found that tweets expressing sympathy and emotional support constituted on average 20\% of the crisis-related datasets. The four crises in which the messages in this category were most prevalent ($>$40\%) all pertained to instantaneous disasters (which included terror attacks). 

\subsection{News Sentiment and Polarity Analysis}

Typically, assessing news articles for polarity involves classifying text for three-valued sentiment: positive, neutral, and negative (e.g., \cite{Reis2015}). However, recently researchers have taken more fine grained approaches towards modeling complex emotions, such as Lin and Margolin's \cite{Ru2014} work on quantifying the diffusion of fear, sympathy, and solidarity during the Boston bombings, and Schulz et al.'s \cite{Schulz2013} work on multi-valued sentiment classification. Moreover, Vargas et al. \cite{Vargas2016} showed that there are marked differences between the overall tweet sentiment and the sentiment expressed towards subjects mentioned in the tweets. Mejova et al. \cite{Mejova2014} took a data-driven approach to understand how controversy interplays with emotional expression and biased language in news using crowdsourcing, and found negative affect and biased language prevalent in controversial issues, while using strong emotions is tempered. 

\section{Methodology}

\subsection{Defining News Sympathy and Sentiment}

To quantify sympathy, we needed firstly to define sympathy. We use Decety \& Chaminade's \cite{Decety2003} definition where ?sympathy is an affective response that?consists of feelings of concern for the distressed or needy other person?. While attempts have been to capture sentiment through tokenization \cite{Pang2008}, both sentiment and sympathy are hard to capture computationally. However, sympathy \cite{Ru2014} and sentiment \cite{Diakopoulos2010} can be detected from text. Therefore, we consider sympathy (is the tweet sympathetic or not towards the affected individuals?) and sentiment (is the tweet negative, neutral, or positive?) as relevant variables to study in higher resolution whether Arab and Western media covered the two attacks differently.

\subsection{Approach Overview}

\begin{figure}[t]
\centering

\scriptsize
        \includegraphics[width=1\columnwidth]{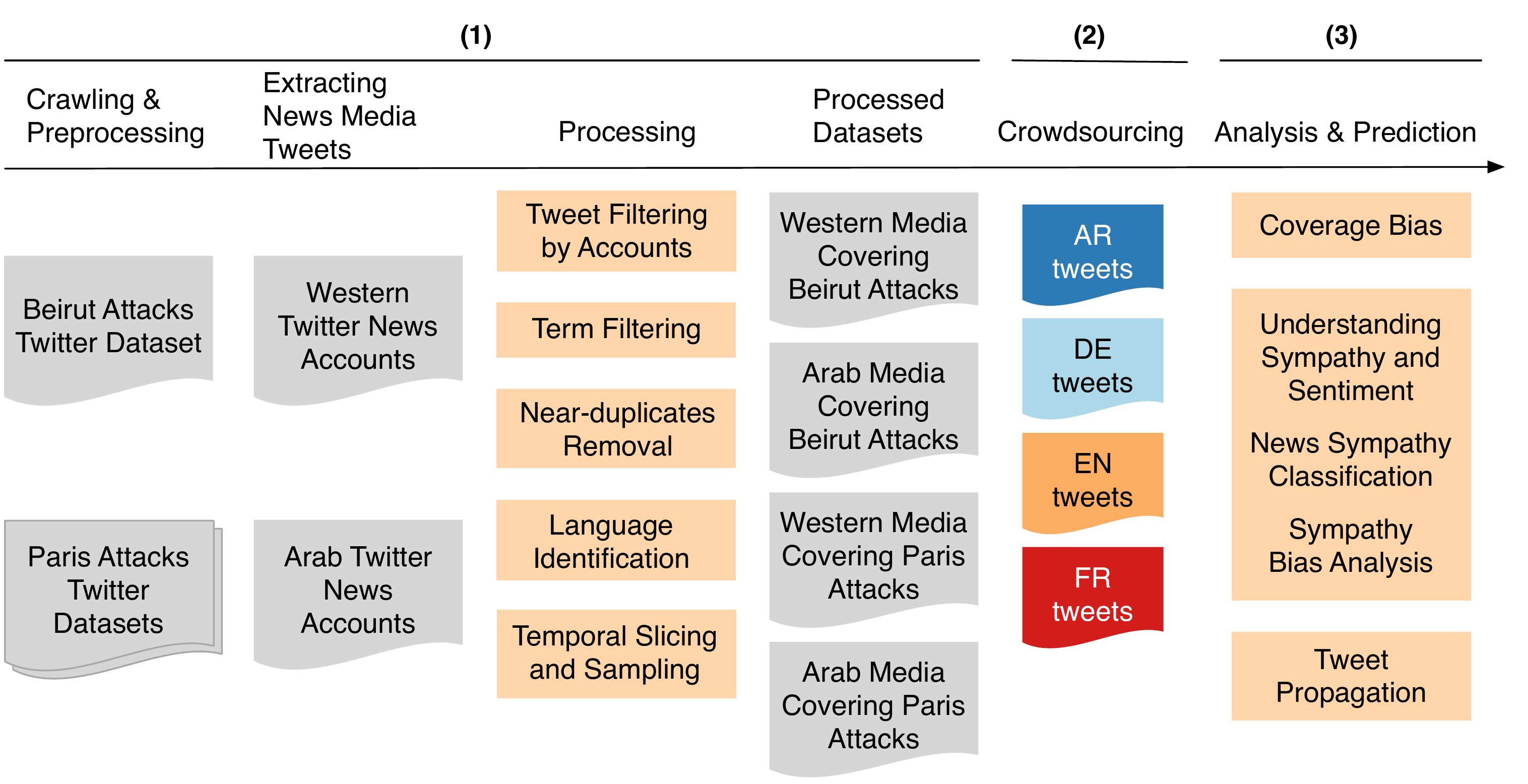}
    \label{fig:a}  
    \vspace{-1pc}

    \caption{Overview of methodological pipeline.}\label{fig:pipeline}

\end{figure}

To build a dataset of tweets from news media accounts for later analysis to answer our research questions, we employ the following methodology (pipeline shown in Fig. \ref{fig:pipeline}): (1) Twitter Data Collection \& Preprocessing (2) Crowdsourcing Annotation (3) Analysis and Prediction. The first two steps are described in detail below, the third in the following section.

\subsection{Twitter Data Collection \& Preprocessing}

We have collected and used four different datasets (described below): Beirut, Paris, Paris Ruest, and Paris Combined.

\textbf{Beirut dataset:} We collected 906,538 tweets on the Beirut bombings on November 12, 2015 shortly after news breakout (approx. 21:00 CET), using the Twitter Streaming API and with these hashtags:  \#lebanon, \#beirut, \#beirut2paris, \#beirutattacks, \#beirutbombing. The dataset was pruned for duplicates. Collection spanned 3.31 days, from 2015-11-12 18:51:07 UTC till 2015-11-16 02:17:04 UTC. Our dataset consisted of 667,073 (73.58\%) retweets and 610,879 unique users. After removing retweets, our dataset had 239,093 unique tweets. The top five hashtags for this dataset are shown in Table \ref{table:hashtags}. What is immediately striking here are the high occurrences of the hashtags \#paris and \#parisattacks, where we attribute this due to the overlap between people's attention to the Paris attacks after having heard about the Beirut attacks. 

\begin{center}  

\begin{table}[h]
\centering
\scriptsize

\begin{tabular}{ c  c | c  c }
\label{table:participants} 
     \textbf{Beirut} && \textbf{Paris (merged)} & \\ \hline
    \textbf{Count} & \textbf{Hashtag} & \textbf{Count} & \textbf{Hashtag}  \\ \Xhline{2\arrayrulewidth}			
    95,280 & \#beirut & 6,777,117 & \#parisattacks \\
    66,898 & \#paris  & 5,765,292 & \#paris \\
    56,021 & \#lebanon & 1,672,152 & \#prayforparis \\
    41,245 & \#parisattacks & 778,132  & \#bataclan \\
    21,546 & \#isis & 760,320 & \#porteouverte \\ \Xhline{2\arrayrulewidth}
\end{tabular}
\caption{Top five most frequent hashtags for each dataset.}
\label{table:hashtags}
\end{table}
\end{center}

\vspace{-1pc}

\textbf{Paris dataset:} We collected a total of 5,339,452 tweets during the two days (13th and 14th) after the November 2015 Paris attacks using the Twitter Streaming API and with the following hashtags: \#paris, \#france, \#parisattacks, \#prayforparis, and \#porteouverte. Some of these JSON records (0.03\%) were poorly structured and others were duplicates, and thus removed. This resulted in a total of 5,337,840 tweets. Collection spanned 1.17 days, from 2015-11-14 13:30:49 UTC till 2015-11-15 18:06:54 UTC. Our Paris dataset had a total of 4,045,046 (75.78\%) retweets and 2,538,348 unique users. 

\textbf{Paris Ruest dataset:} In addition to our collection, we use a larger dataset created by Nick Ruest\footnote{http://ruebot.net/post/look-14939154-paris-bataclan-parisattacks-porteouverte-tweets; last retrieved: 24.12.2017}, who collected tweets shortly after the attacks occurred (approx. 23:00 UTC on November 13) with the following hashtags: \#paris, \#parisattacks, \#prayforparis, and \#porteouverte. We hydrated (collected metadata for tweets based on tweet IDs) the dataset on 2015-11-19 using the Twitter public API, and collected a total of 12,788,201 tweets. Tweet volume was less (attrition rate: 14.4\%) than the original collection (N=14,939,154), which is common given that some tweets are removed (either by Twitter or by users). This collection spanned 33.96 days, from 2015-11-04 21:14:39 UTC till 2015-12-08 20:54:03 UTC. NH's Paris dataset had a total of 9,742,241 (76.18\%) retweets and 4,127,762 unique users. 

\textbf{Paris Combined dataset:} For later analysis, we combined both datasets. We expanded the Paris dataset (by merging with the Paris Ruest dataset) because it spanned only 1.17 days after the Paris attacks, while the Beirut dataset spanned 3.31 days. This step is important for temporally aligning the Paris dataset with the Beirut dataset. Additionally, the imbalanced dataset size notwithstanding, we later downsample all datasets to ensure the datasets are balanced and any effects seen are not an artifact of a large sample size. Merging datasets resulted in a total of 16,868,318 tweets. Our dataset consisted of 75.6\% retweets, resulting in a total of 4,110,291 unique tweets with 4,013,489 unique users. This combined Paris dataset (with retweets and duplicates removed) is used for all subsequent analyses, and will be referred to simply as the Paris dataset. 

\subsection{Crawling \& Extracting News Media Tweets}

To answer our questions about media coverage bias (specifically differences in news sympathy), we apply multiple processing steps. We first crawl Twitter news (and blog) accounts, extract the news tweets from our datasets, identify the language of the tweets, slice our data by time to ensure that the two event timelines match in duration, and finally draw samples to ensure our data is sizeable and ready for human computation. Details of each step is described below.

\begin{center}  

\begin{table}[t]
 \centering
\scriptsize
\begin{tabular}{ c  | p{5.3cm} }
\label{table:participants} 
    \textbf{Region} & \textbf{Country (population per million)}  \\ \Xhline{2\arrayrulewidth}			
    Middle East (6) & Egypt (89.6), Iraq (34.8), Saudi Arabia (30.9), United Arab Emirates (9.1), Jordan (6.6), Lebanon (4.5)
 \\ \hline
    Western (5) & USA (318.9), Germany (80.9), France (66.2), United Kingdom (64.5), Spain (46.4)
 \\ \Xhline{2\arrayrulewidth}
\end{tabular}

\caption{Regions and countries of interest for our analysis. Population estimates drawn from the World Bank (http://www.worldbank.org/; last retrieved: 24.12.2017)}
\label{table:countries}
\end{table}
\end{center}

\vspace{-1pc}

\subsubsection{Crawling Twitter news accounts}

Our first step was to identify and collect influential Western and Arab news accounts on Twitter. The list of countries chosen across the Middle East and the Western world are shown in Table \ref{table:countries}. With respect to the Middle East, we chose countries that were geographically near Lebanon, and that did not have explicit and/or visible political nor religious conflicts with Lebanon at the time of collection (e.g., Syria was excluded due to the ongoing conflict at the time of data collection). For Western countries, we based our decision on population size, language (English being most prominent), and proximity to France. Despite that Twitter is dominated by English language users\footnote{http://www.beevolve.com/twitter-statistics/\#a3; last retrieved: 24.12.2017}, we wanted to ensure that we were collecting news media tweets from both English as well as the native language of the countries of interest. To find news media\footnote{We use Wikipedia's definition of news media, which includes blog accounts: https://en.wikipedia.org/wiki/News\_media ; last retrieved: 24.12.2017} accounts on Twitter from these countries, we followed a two-step approach: 

We found a seed set of accounts automatically (using Twitter's relevance-based Search API) by crawling user accounts (see Supplementary Material B) with news related queries (e.g., `France news' for English queries; `Nouvelles France' for native language queries). This first step deliberately takes a crude computational approach as curating news organizations by experts may be subject to bias, and could exclude unfamiliar news accounts that possibly became highly active during the time of crisis (e.g., bloggers new to the scene). We included bloggers because (a) Bloggers play an important role in disseminating news, cf., during the Arab spring in Tunisia and Egypt \cite{Lotan2011}) (b) They are typically uncensored, since they do not `officially' provide public service \cite{Singer2006}.

To ensure a measure of influence, we only retrieve user accounts with at least 5k followers. Despite earlier research that showed a high number of followers does not always mean an influential user \cite{Cha10}, we used follower count as a simple heuristic to gather prominent news accounts. While our collection could also have include state-sponsored news (which is difficult to identify), our intention was to capture tweet samples that users would likely see. We also checked for romanized Arabic queries (e.g., Akhbar Masr), but did not find additional accounts matching the >5k follower requirement. Our queries returned six results (which we kept) with a follower count <5k (Max=4,935, Min=4,473), with the rest >5k (Max=31.4M, ``\textit{CNN Breaking News}"). We did not set a limit on account creation date nor on Twitter verification, as our experiments showed: (a) new bloggers and news agencies with accounts created only a year earlier (2014) appeared to be quite active in reporting events (b) even major news accounts were sometimes not Twitter verified, wherein we could potentially miss important news accounts if we enabled this filter. 

Returned accounts were manually inspected to ensure they comprise news media outlets and blogger accounts. As a sanity check, we also cross-checked whether account names occur in public lists (e.g., Wikipedia pages `News media in \{Country\}'\footnote{E.g., https://en.wikipedia.org/wiki/News\_media\_in\_the\_United\_States; last retrieved: 24.12.2017}) and for blogger accounts, whether they cross-link to a webpage. It is important to mention here that there is no such benchmark list of news media accounts on Twitter. We had a total of unique 208 news media accounts, where 93.3\% (194/208) of our dataset consists of news outlet accounts, and the remainder 8 journalist and 6 blog accounts. Furthermore, there was some overlap in accounts for Western media (38/117) coverage and for Arab media (38/91) coverage of Paris and Beirut. The final list of unique crawled news organizations (N=208) is provided as a supplementary dataset to this paper (see Supplementary Material C)\footnote{We show the name, user name, user description, the country / query used to retrieve the account, and follower count at the time of crawling -- all of which are publicly available data. Additionally, we include tweet count and mean scores and standard deviations across all labels.}. 

\vspace{-1pc}

\subsubsection{Tweet Filtering by Accounts}

After gathering a set of news accounts across countries using both English and native language queries, we then matched these user IDs with all IDs in our Beirut and Paris datasets. The full set of queries used, the total number of Twitter news accounts found, and the amount and percentage extracted from both datasets are shown in Supplementary Material B. This process resulted in four datasets: (1) Arab media covering the Beirut attacks (N=2,766) (2) Arab media covering the Paris attacks (N=2,728) (3) Western media covering the Beirut attacks (N=245) (4) Western media covering the Paris attacks (N=9,245). The datasets combined resulted in 14,984 tweets.

\vspace{-1pc}
\subsubsection{Term Filtering}

As an additional step, we made sure that within each dataset, there was no mention of the other set of attacks (e.g., we removed all mentions of Paris from the Beirut dataset), and that all tweets pertained to the events in question. Even though the two attacks happened a day apart, where we would expect cross-pollination across messages, we deliberately chose not to include tweets that reference both the Paris and Beirut attacks, as this may influence our attempts at investigating media bias within each tweet dataset separately. For the Beirut dataset, we filtered out tweets that included these terms: paris, par\'{i}s france, parisattacks, bataclan, parisattacks, porteouverte. For the Paris dataset, we filtered out the following terms: beirut, lebanon, beirutattacks, \RL{\small libnAn} [Lebanon], \RL{\small bayrwt} [Beirut]. 

\vspace{-1pc}
\subsubsection{Near-duplicates Removal}

Finally, to ensure that our dataset contains only unique tweets without any near duplicates (as this would cause redundancy later in the annotation task), we removed all partial duplicates from our resulting datasets. This was done by applying the Levenshtein distance \cite{Levenshtein1966} string similarity metric on the tweet texts of each dataset, with a threshold set to 0.1. This reduced the size of our dataset to 12,814 tweets.

\vspace{-1pc}

\subsubsection{Language Identification}

To prepare our Beirut and Paris datasets for analysis of sympathy, we need to be able to identify the language of the tweets so crowdworkers can annotate them. To do so, we used the langid.py \cite{Lui2012} language identification Python package, and computed the (percentage) distribution of languages. We used langid.py as it provided us with classification probabilities while Twitter's `lang' value does not provide such a metric, so we could manually adjust the `lang' of low confidence tweets. To deal with any misclassifications from langid.py, we disqualified any tweets with a normalized classification probability of $<$ 0.95, reducing our dataset to 10,460 unique tweets. For the remainder of the tweets, we manually inspected and reclassified all tweets with a normalized probability of $>$ 0.95. We found that across all datasets combined, the languages of tweets were either English (67\%), German (13.9\%), Arabic (10.1\%), French (8.3\%), or Spanish (0.7\%). Given the low percentage of Spanish in our dataset, we decided to exclude any Spanish tweets from all subsequent analyses. 

\subsubsection{Temporal event slicing and sampling}

Given that our Beirut and Paris datasets differed temporally in coverage of the attacks, we need to normalize coverage duration as tweets posted 5 days after the attacks may differ for example than tweets posted two weeks after the attacks. Since we were constrained by the size and coverage duration of our Beirut dataset, we used the coverage length of that dataset as a seed to slice the Paris dataset. For our processed Beirut dataset, our earliest coverage started on 2015-11-12 18:52:30 UTC (approx. 4 hours after the Beirut attacks that took place around 18:00 EET) and went until 2015-11-16 02:00:10 UTC. This amounts to exactly 3.3 days. Thereafter, we applied the same time slice for the processed Paris dataset, where earliest coverage started from 2015-11-13 21:15:20 UTC (approximately one hour after the Paris attacks, which started at 20:20 UTC) until 2015-11-17 02:30:23 UTC, giving exactly 3.22 days. This time slicing further reduced the total size of our combined datasets to 7,768 unique tweets: Western media coverage of Beirut (N=131), Western media coverage of Paris (N=5,298), Arab media coverage of Beirut (N=287), and Arab media coverage of Paris (N=1,566).

\begin{figure}[t]
\centering
\scriptsize
        \includegraphics[width=0.8\columnwidth]{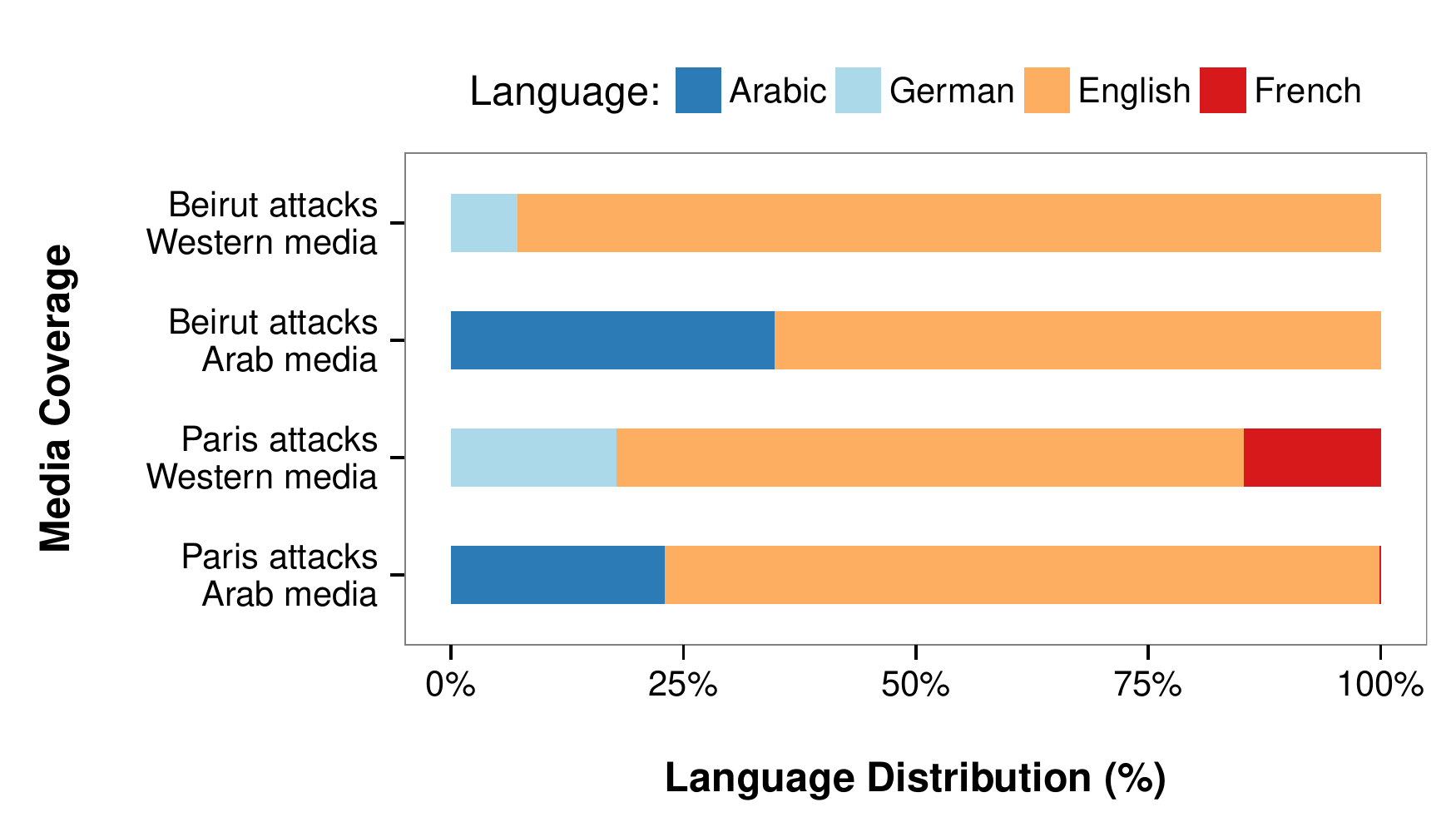}
    \label{fig:a}  
    \vspace{-1pc}
    \small \caption{Language distribution of tweets to be annotated across datasets.}\label{fig:langdist}
\end{figure}

Finally, to send our tweets for annotation by crowdworkers, we only needed a sufficient sample from each of our four resulting datasets to avoid lengthy crowdwork time and costs. Even though this annotated data is limited, we later train and classify our unlabeled data automatically which handles the limitations of crowdsourcing. Therefore, we drew a random sample of 1,000 tweets from the Paris datasets. However, random sampling may miss important tweets that occurred on specific days within our 3.3 days. Therefore, we split each dataset into separate buckets of approximately 24 hours, and drew normalized random samples from each bucket to eliminate bias in drawing more samples from a day that happens to have more records. The normalization constant was calculated by dividing the size of the desired sample draw (1,000) by the total number of rows in each dataset. For each bucket, the sample drawn was the number of records in that bucket multiplied by the normalization constant, and rounded to ensure all day buckets cap at 1,000 records. This process reduced the size of the Paris attacks datasets (Western and Arab media coverage) each to N=1,000. The final language distribution of our language-specific datasets ready for annotation is shown in Fig. \ref{fig:langdist}. 

\subsection{Crowdsourced Overall Sympathy Annotation}

To annotate our datasets, we employed crowdworkers through the CrowdFlower\footnote{http://crowdflower.com/; last retrieved: 24.12.2017} platform. We had four language-specific datasets with corresponding annotation instructions. A language-specific dataset covered both the Beirut and Paris attacks, which meant there was an assumption that a worker had to be familiar with both events to accurately label tweets. For each language-specific annotation task, we provided task instructions and examples in that target language (translated from English by native speakers). For each tweet (and any accompanying media), a worker had to label it for sympathy (sympathetic\footnote{E.g., thoughts, prayers, sadness, solidarity with affected individuals.}, unsympathetic) and sentiment (positive, neutral, negative), and optionally `NA'  for non-applicable tweets.

For worker selection, we ensured workers spoke the target language. We did not set a restriction on worker location, as a worker could speak a target language (e.g., Arabic) yet reside in a non-Arabic speaking country. While we set target language requirements, we did not make our tweets into images (cf., \citeyear{Bergsma2012}), which can be a limitation. Following standard guidelines from CrowdFlower, 10-15 tweets per language-specific task were classified by the paper authors. We did not trust assessments of any worker who differed significantly from our own (cut-off point of $<$ 70\% agreement). Workers were presented with the original tweet, and included media items (image or video), and were asked to label it. While we are aware of the potential ethical concerns on behalf of Twitter users in displaying their name (cf., \citeyear{Olteanu2015}), in our case we were only displaying tweets from news organizations, who are presumably aware and even encourage publicizing their content. Importantly, omitting the username of our tweets would risk misrepresenting the original tweet and its overall sympathy. 

Trusted workers took on average (across all languages) 57 seconds (interquartile mean) to label each tweet. We collected labels from at least 3 different trusted workers per tweet and task\footnote{Note that if a tweet has different labels from all 3 workers, the CrowdFlower platform brings in additional workers.}, where the final label of the tweet was determined by majority vote. We followed the guidelines of CrowdFlower, and set a limit of no worker labeling more than 300 items in our rating task. Workers were paid 10 cents per page, where each page contained 5 tweets. This amounted to approximately \$10 per 100 tweets. In the end, we had a total of 2,390 x 2 x 3 =  14,340 labels (excluding the `not applicable' checkbox).

\begin{center}  
\begin{table}[t]
 \centering
\scriptsize
\begin{tabular}{ l  c l  c  c  c}
\label{table:participants} \\ \Xhline{2\arrayrulewidth} 
   \textbf{Lang.} & \textbf{\# Tweets} & \textbf{Label} & \textbf{$a$=3 (\%)} & \textbf{Fleiss' Kappa} \\ \Xhline{2\arrayrulewidth}         
	EN & 1732 & Sentiment & 69.3 & 0.27 \\
          &    &   Sympathy & 83.4 & 0.42 \\
           \hline
	AR & 354 & Sentiment & 69.1 & 0.32 \\
          &    &   Sympathy & 82.7 & 0.29 \\  \hline
	FR & 147 & Sentiment & 73.1 & 0.22 \\
          &    &   Sympathy & 82.4 & 0.32\\ \hline
	DE & 185 & Sentiment & 73.1 & 0.28 \\
          &    &   Sympathy & 85.5 & 0.38\\ \Xhline{2\arrayrulewidth}
       
\end{tabular}
\caption{Trusted worker agreement ($a$=3) scores across languages (from CrowdFlower) and our own computed Fleiss' Kappa scores.}
\label{table:crowdworkers}
\end{table}
\end{center}

\vspace{-1.5pc}

\subsubsection{Annotation Quality}

Despite that labeling sentiment can be subjective (cf., \citeyear{Pang2008}), we provided detailed instructions and examples of Positive, Negative, and Neutral tweets (see Supplementary Material D) to ensure workers correctly label the data. Moreover, our inter-rater agreement scores drawn from CrowdFlower (Table \ref{table:crowdworkers}) are promising, with the lowest being 69.1\% for sentiment classification of Arabic tweets, which is in line with previous work, and highlights the reliability of crowdsourced social media annotation \cite{Olteanu2015,Diakopoulos2012}. As a further measure, we also computed Fleiss' Kappa for the labeled tweets, and found reasonable agreement scores (Table \ref{table:crowdworkers}), with sentiment expectedly exhibiting lowest agreement across languages\footnote{Since sentiment is 3-valued instead of binary, it could more likely result in higher rating variance.}. Furthermore, we followed Olteanu et al.'s \shortcite{Olteanu2015} approach and independently (N=2) rated a random sample of 15 tweets from each language dataset\footnote{We ensured that the tweets were translated by native speakers.} (total N=60) and computed unweighted Cohen's Kappa for each factor except sentiment (which was weighted). Our ratings reached substantial agreement on sentiment ($\kappa$=0.70, CI: [0.52,0.88]) and sympathy ($\kappa$=0.71, CI: [0.50,0.91]) labels. Thereafter, we took our agreement ratings and compared their joint label with those provided by workers, and we reached reasonable agreement for sentiment ($\kappa$=0.64, CI: [0.42,0.87]) and sympathy ($\kappa$=0.59, CI: [0.37,0.82]). 

Access to our anonymized data set is available (see Supplementary Material A) and on GitHub: \url{https://github.com/abdoelali/CrisisNewsSympathy}. 

\section{Results}

\subsection{Coverage Bias}

\begin{figure}[t!]
\centering
\begin{subfigure}{.23\textwidth}
  \centering
  \includegraphics[width=0.9\linewidth]{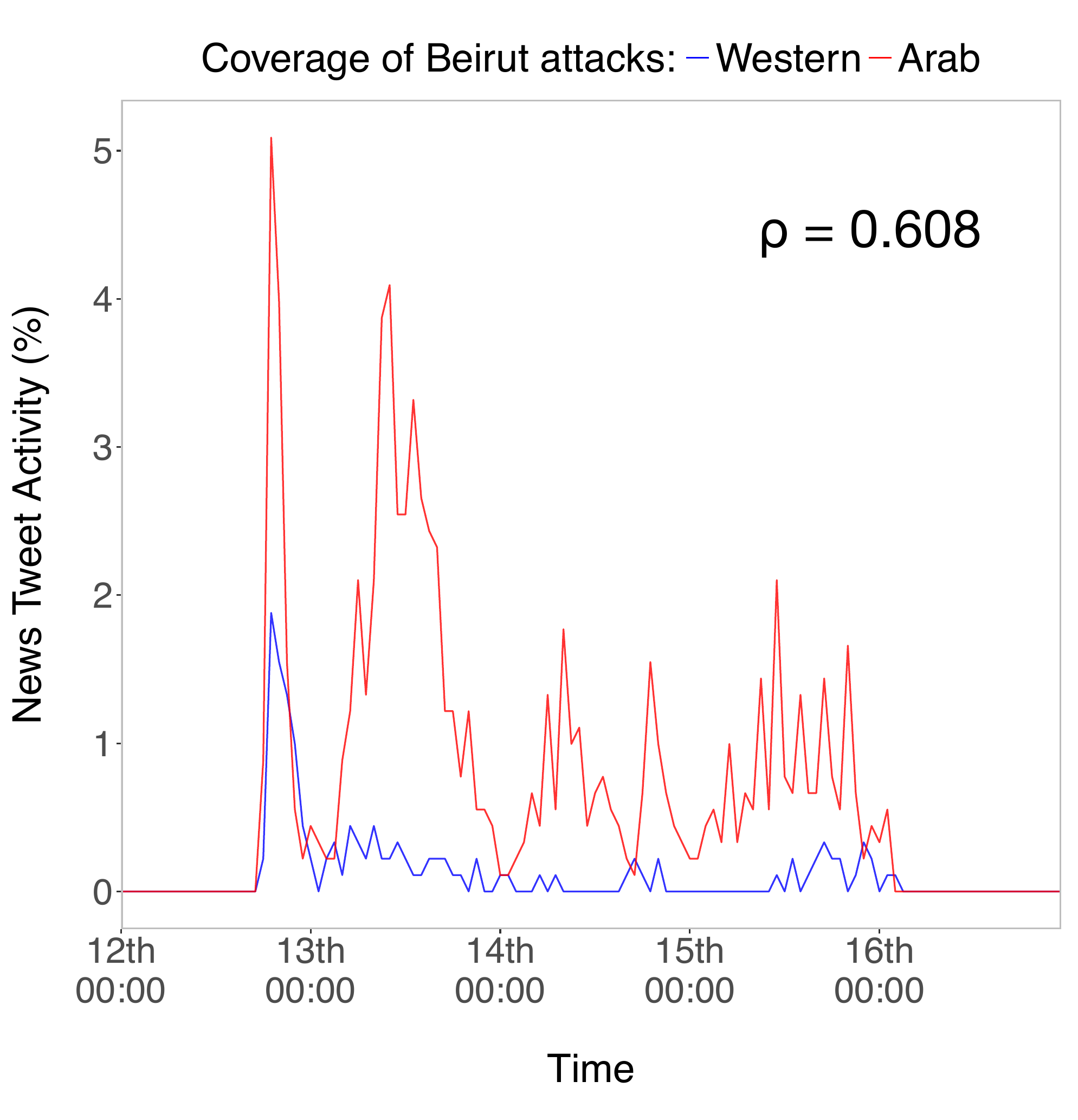}
  \label{fig:act_time_lb}
\end{subfigure}%
\begin{subfigure}{.23\textwidth}
  \centering
  \includegraphics[width=0.9\linewidth]{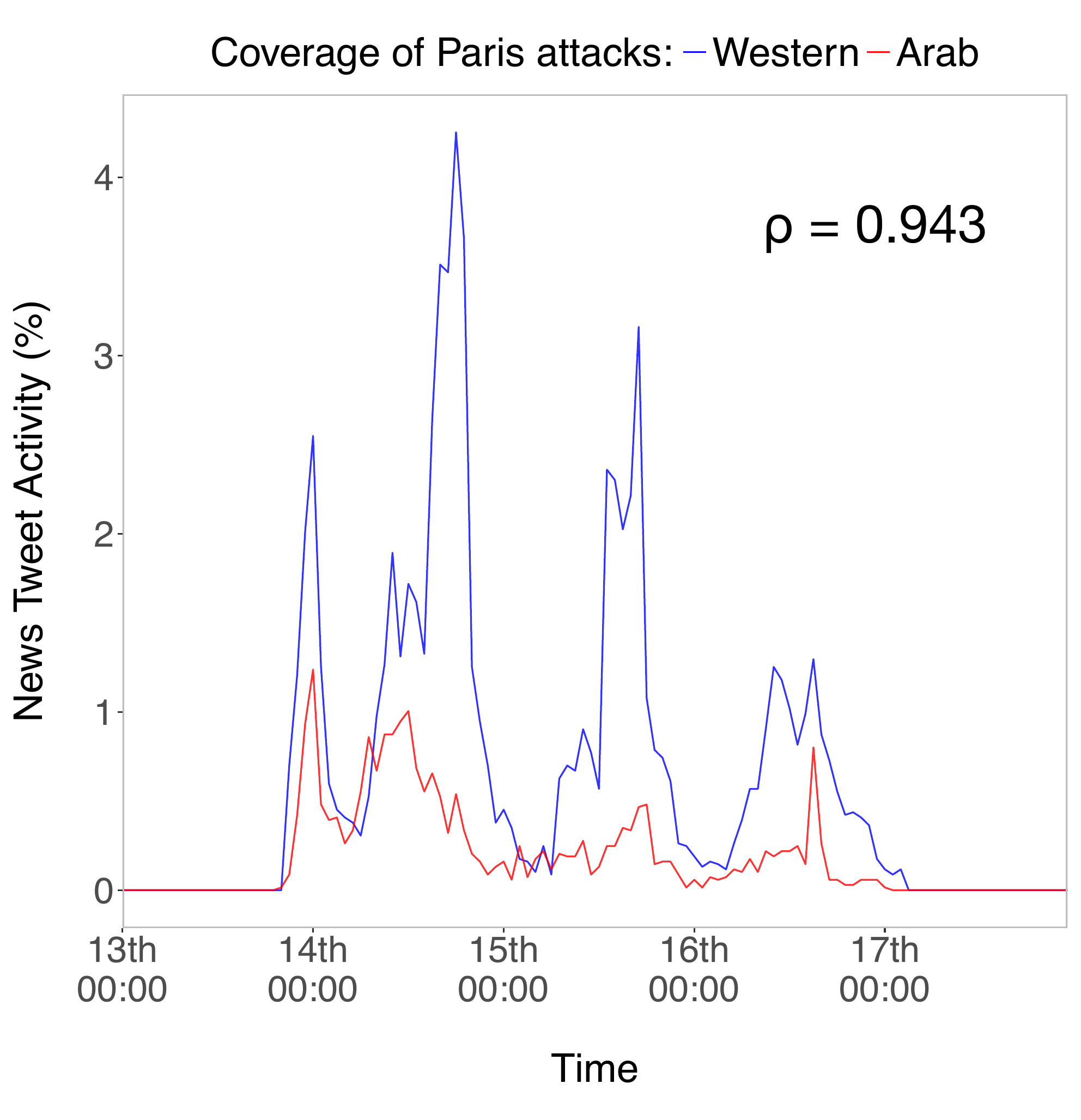}
  \label{fig:act_time_fr}
\end{subfigure}
\vspace{-1pc}
\caption{Hour by hour normalized tweet activity volume across the days after the Beirut attacks (left) and Paris attacks (right). \textit{Best seen in color.}}
\label{fig:act}
\end{figure}

We looked at normalized tweet volume from Western and Arab media across both attacks (N=7,768), and visualize this daily and hourly in Fig. \ref{fig:act}. Attention bursts reflect day-night cycles. From the graphs, we can see that for the Beirut attacks, there was more coverage from Arab media, with the inverse for the Paris attacks, which shows more Western media coverage. We ran a Chi-square test with Yates' continuity correction across all days to compare the difference between Arab and Western media coverage. In line with our hypothesis, our test revealed a statistically significant difference in tweet activity volume between how much Western ($M$=0.909) and Arab media ($M$=5.37) covered the Beirut attacks and by how much Western ($M$=36.79) and Arab media ($M$=10.87) covered the Paris attacks ($\chi^2$(1, N=7,768) = 1489, p$<$0.001, $\phi$=0.44, odds ratio=0.05). From this, we accept the alternative hypothesis that there was indeed coverage bias across both attacks.

Furthermore, did Western and Arab media follow a similar pattern of tweeting? We ran correlation analyses between hourly tweet volume for both attacks across media coverage, and found that for the Beirut attacks, Western and Arab media exhibited a medium-sized correlation (Spearman ($\rho$) = 0.608, p$<$0.001; $\phi_c$=0.698) and for the Paris attacks exhibited a significant and strong correlation (Spearman ($\rho$) = 0.943, p$<$0.001; $\phi_c$=0.918). We see that for tweet activity volume, Western and Arab media are engaged at approximately similar time points, which supports the fairness of our collected data on Western and Arab media across these two attacks.

\subsection{News Media Sympathy Bias}

\subsubsection{Understanding Sympathy and Sentiment}

\begin{figure}[t]
\centering
\scriptsize
\begin{subfigure}{0.2\textwidth}
  \centering
  \includegraphics[width=0.9\linewidth]{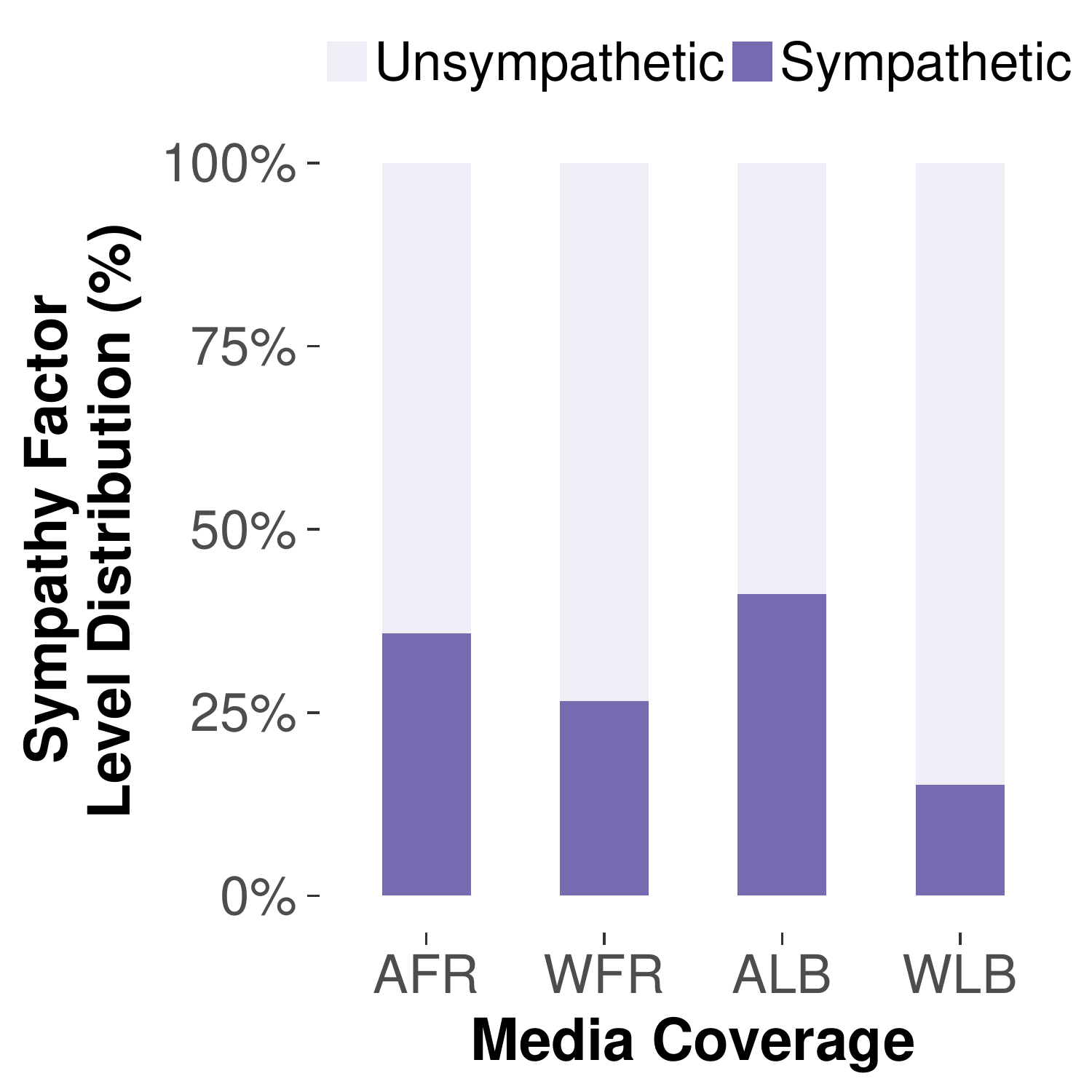}
    \subcaption{Sympathy}

\end{subfigure}
\begin{subfigure}{0.2\textwidth}
  \centering
  \includegraphics[width=0.9\linewidth]{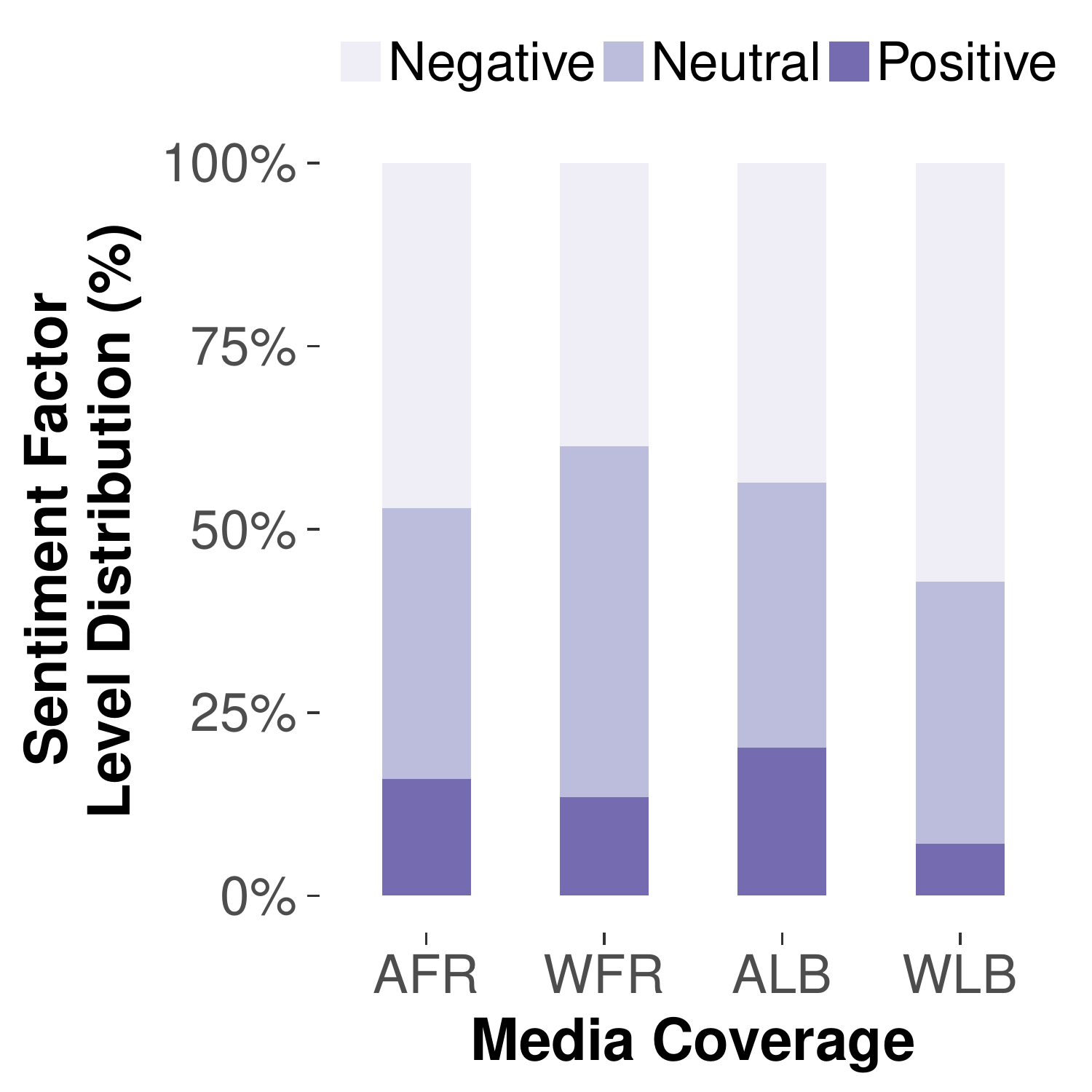}
  \subcaption{Sentiment}
\end{subfigure}

\caption{Factor level distributions for Arab media coverage of Paris attacks (AFR), Western media coverage of Paris attacks (WFR), Arab media coverage of Beirut attacks (ALB), Western media coverage of Beirut attacks (WLB).}
\label{fig:prop_dist}
\end{figure}

The percentage level distributions for sympathy and sentiment are shown in Fig. \ref{fig:prop_dist}. As previous work showed sentiment alone may not be enough to capture subtleties, especially in crisis data \cite{Olteanu2015,Schulz2013,Vargas2016}, we needed to build on this prior work and explore whether sentiment may be a useful construct in our analysis. To understand the relationship between sentiment and sympathy, we developed a binomial logistic regression model, since the categorical variable sympathy to predict is binary. We report our results in Table \ref{table:logit}. We did this to delve deeper into the relationship between these two variables. No multicollinearity effects were found. Our regression model (Eq. \ref{eq:regression}) is as follows:

\vspace{-1pc}
\begin{equation}
\small
\begin{split}
\text{logit}(p_{Symp})= \log\left(\frac{p_{Symp}}{1 - p_{Symp}}\right) = \beta_0 + \beta_1 S_i 
\label{eq:regression}
\end{split}
\end{equation}

where: $p_{Symp}$ is the probability that a tweet is sympathetic, given the predictor sentiment ($S_i$). $p_{Symp}$ ranges between 0 and 1, so $p_{Symp} \in \{0,\dotsc,1\}$. $\beta_0$ is the intercept from the linear regression equation (the value of the criterion when the predictor is equal to zero). $\beta_1 S_i$ is the regression coefficient multiplied by our sentiment predictor.

We first present the goodness of fit of our model and how it fares against the (intercept) null model (Table \ref{table:logit}). Compared to the null model, our model provides considerable explanatory power with significant reduction in deviance. However, while the difference may be statistically significant, the difference between model deviances do not differ much, as is evident by the goodness of fit McFadden $R^2$, which is 0.2. Testing our sentiment\footnote{Sentiment is split between neutral and positive in our models due to dummy variable coding for non-binary categorical variables.} model against the null model yields significant results ($\chi^2$(1, N=2,390) = 2425.7, p$<$0.001). To test out the performance of our model, we used 80\% of our data for training purposes and parameter tuning and the remaining 20\% were held out for testing. For training, we used $k$-fold ($k$=10) cross-validation given the small dataset size (N=2,390). Next we report our results on the 20\% held-out test dataset. Accuracy of our model is 79\% (Precision = 0.79, Recall = 0.41, F-1 = 0.54), which highlights that sentiment appears to have sufficient power for predicting sympathy. Nevertheless, while sentiment is important, from our data it is also clear that is different from sympathy. 

To better understand the relationship between sympathy and sentiment, we further tested correlations\footnote{We use Spearman's rho in addition to Cramer's V (since our data is not normally distributed) by making an assumption that our dichotomous variables exhibit a monotonic relationship, and therefore can be treated as ordinal variables. Moreover, Cramer's V is symmetric, and does not show negative relationships (i.e., no correlation direction).} between them. There was a strong positive correlation between sentiment and sympathy (Spearman ($\rho$) = 0.19, p$<$.001; Cramer's V ($\phi_c$) = 0.49), indicating we are likely measuring closely related concepts. These results support our regression model coefficients. Given the results of of both our regression modeling and correlation analyses, we find that sympathy as a concept is related to sentiment. Since our crowd sourcing results analysis revealed that they are more or less capture similar signals, we chose to focus on sympathy. Below, we present our approach to predict sympathy from the tweet text, where we investigate sympathy classification performance on unseen (i.e., unlabeled) data.

\subsubsection{Classifying News Media Sympathy}
\label{sec:classifier}

The crowdsourced annotation task was not able to cover the entire corpus due to practical constraints (e.g., the cost and availability of workers). To generalize the analysis to unlabeled data, we also developed machine learning models that recognize the sympathy of tweets. The learning task was designed as a binary classification problem where a model aims to classify if a given tweet is sympathetic or unsympathetic. A classification model was built for each language, and each model was trained using the data annotated through crowdsourcing and applied to the rest. For the classification model, we adopted a convolutional neural network (CNN) combined with word2vec embeddings (cf., \cite{Kim2014}). This work extended the recent successful applications of deep learning models in NLP tasks to sentence classification and showed the model achieves high performance even with a very simple architecture (i.e., one convolutional layer) and little tuning, which keeps parameter count low and reduces needed training samples. This CNN architecture has a single channel (word2vec embeddings) and three layers: the embedding layer which translates words of a sentence to corresponding word2vec embeddings; convolutional layer that applies filters over sliding windows of words and extracts a feature through max-pooling; final layer performs dropout regularization \cite{Srivastava2014} and classifies the result using softmax (see \cite{Kim2014} for more details of the model).

\begin{table}[t]
\centering
\scriptsize
{\begin{tabular}{ll}
 &  \multicolumn{1}{c}{\textbf{Characterizing Sympathy}} \\       \midrule\\[-5.5ex] \\
  & \textbf{Model} \\ 
      \midrule\\[-5.5ex] \\
  \textbf{Variable} & \multicolumn{1}{c}{\textbf{Coefficient (Standard error)}} \\
  \hline
Sentiment$_{NEUT}$ &  -1.0*** (0.07) \\ 
Sentiment$_{POS}$ & \phantom{-}2.38*** (0.15) \\ 
    Intercept & -0.79*** (0.07) \\ 
  McFadden's $R^2$ & \phantom{-}0.2\\
 \midrule\\[-3ex]
   Deviance &  \phantom{-}2453.6\\
    AIC & \phantom{-}2453.6\\
    \midrule\\[-2.5ex]
   \multicolumn{2}{l}{*p$<$.05, **p$<$.01, ***p$<$.001}  \\
    \midrule\\[-2ex]
\end{tabular}}
\vspace{-1pc}
\caption{Logistic regression model examining sentiment factor associated with tweet sympathy.}
\label{table:logit}
\end{table}

Since we pre-trained embeddings on our entire corpus, it makes binary classification easier to fit on smaller data.  Also, complexity of conventional models (logit regression, SVM) is not low for NLP tasks: n-grams easily make the feature space very large. They often require language specific pre-processing and feature selection techniques, which increases complexity since we deal with multiple languages. Instead, such problems (e.g., capturing local context \cite{Goldberg2015}) are dealt with by the CNN model itself, increasing clarity and method replicability. The word2vec embeddings were pre-trained for each language \cite{Mikolov2013}. Given a text corpus, word2vec learns a lower-dimensional vector (100 dimensions in our experiment) representation of words that preserves the semantic distance between them. As we deal with tweets for a particular topic, we used the entire set of collected tweets for training the embeddings instead of using an available general corpus. Since we pre-train on our entire corpus, it makes binary classification easier to fit on smaller data.  Also, complexity of conventional models (e.g., SVMs) is not low for NLP tasks: n-grams easily make the feature space very large. Especially since we deal with multiple languages. Instead, such problems (e.g., capturing local context \cite{Goldberg2015}) are dealt with by the CNN model itself, increasing clarity and method replicability.

We briefly describe the configuration of the model (the details can be found in the TensorFlow implementation we used \cite{Britz2017}). The model is trained through stochastic gradient descent with the Adadelta update rule \cite{Zeiler2012}. It trains for 100 epochs using shuffled mini-batches of 64 instances. Three different filter sizes (3, 4, and 5 words) were used for convolution, and 128 filters were made for each size. Dropout rate was set to .5. For evaluation of the approach, we ran 10-fold cross validation (Table \ref{table:classify_lang}). We chose balanced accuracy\footnote{Arithmetic mean of class-specific accuracies.} to account for class imbalances in our data, and report the precision and recall for each class. Overall, the weighted average of the balanced accuracies across languages was 72.5\% (shown in Table \ref{table:classify_lang}).

\begin{center}  

\begin{table}[t]
\centering
\scriptsize

{\begin{tabular}{ c  c | c }
\label{table:participants} 

      &  \multicolumn{2}{ c }{\textbf{Overall balanced accuracy: 0.725}} \\ \hline
     & \multicolumn{2}{ c }{\textbf{EN balanced accuracy: 0.762}} \\ \hline
     & Sympathetic & Unsympathetic \\ \hline
     Precision & 0.643 & 0.8882 \\ 
     Recall & 0.568 & 0.911 \\ \hline
      & \multicolumn{2}{ c }{\textbf{AR balanced accuracy: 0.699}} \\ \hline
     & Sympathetic & Unsympathetic \\ \hline
     Precision & 0.65 & 0.748 \\ 
     Recall & 0.381 & 0.846 \\ \hline
     & \multicolumn{2}{ c }{\textbf{FR balanced accuracy: 0.812}} \\ \hline
     & Sympathetic & Unsympathetic \\ \hline
     Precision & 0.783 & 0.841 \\ 
     Recall & 0.32 & 0.964 \\ \hline
     & \multicolumn{2}{ c }{\textbf{DE balanced accuracy: 0.501}} \\ \hline
     & Sympathetic & Unsympathetic \\ \hline
     Precision & 0.867 & 0.135 \\ 
     Recall & 0.811 & 0.208 \\ \hline

\end{tabular}}
\caption{Balanced accuracies (Sympathetic, Unsympathetic) for unlabeled news media tweets (N=5,378) across each language.}
\label{table:classify_lang}
\end{table}
\end{center}

\vspace{-1pc}
\subsubsection{News Media Sympathy Analysis}

Since our sympathy labels are not normally distributed, we ran Mann-Whitney U tests to compare the difference in sympathy between Arab and Western media for each of the datasets on the Paris and Beirut attacks. We found a statistically significant difference in sympathy scores between Western and Arab  media in coverage of the Beirut attacks (Z=5.146, p$<$0.001, r=0.22), as well as between Western and Arab media in coverage of the Paris attacks (Z=4.151 p$<$0.01, r=0.09). To ensure that the effects we observe are not due to imbalances in the size of the compared datasets, we down-sampled all datasets to the minimum dataset size, namely Western media reporting of Beirut (N=112). We used random downsampling without replacement to align each pair of datasets, and to ensure any downsampling biases are removed, we tested differences (Mann-Whitney U tests) in sympathy scores across 1000 sampling runs (seeds)\footnote{Since p-value combination under Fisher's method follows a $X^2$-square distribution, we needed a minimum of 220 runs to achieve 0,95 power and 0.3 effect size under $\alpha$=0.05.} To combine probabilities, we used Fisher's method\footnote{This is a common method used for aggregating probabilities, however we tested other methods (e.g., voting) and results did not differ.} \cite{Fisher1925,Kost2002} shown in Eq. \ref{eq:fisher} below:

\begin{equation}
\small
-2\sum_{i=1}^{k}\log(p_i) \sim x_{2k}^{2} 
\label{eq:fisher}
\end{equation}

where $p_i$ is the p-value for the $i^{th}$ hypothesis test and $k$ is number of sampling runs. When the p-values tend to be small, the test statistic $X^2$ will be large, which suggests that the null hypotheses are not true for every test. Here again, we found a statistically significant differences in sympathy scores between Western and Arab media coverage of the Beirut attacks ($\chi^2$(1, N=2,000) = 22467.62, p$<$0.001) as well as coverage of the Paris attacks ($\chi^2$(1, N=2,000) = 19458.9, p$<$0.001).

Table \ref{table:ratio_positive} presents CNN-based classification results of the unlabeled data, showing the ratio of sympathetic tweets. For Western media coverage of the Beirut attacks, we had very few unlabeled tweets (N=1). Thus we instead report the ratio from the labeled (ground truth) data. The findings described above were consistently observed in the results. Western media had less sympathetic tweets than Arab media in coverage of the Beirut attacks, and the difference in the amount of sympathetic tweets is larger than that for the Paris attacks. Another way to view this is that each media coverage (Arab, Western) was overall more sympathetic towards the country affected in their respective region. Western media was more sympathetic towards Paris, while Arab media was more sympathetic towards Beirut. This aligns with prior work showing strong regionalism in news geography \cite{KwakA2014} and with producer-consumer attention asymmetries across countries \cite{Abbar2015}.

\vspace{-0.5pc}

\subsection{Sympathy and Sentiment in Tweet Propagation}

Finally, we tested whether news sympathy has network effects, specifically whether sympathetic tweets result in higher information propagation. Given prior work on what makes online content go viral (cf., emotionality and positivity in Berger et al.'s work \cite{Berger2012}), we expected sympathy and positive sentiment to be good indicators. Therefore, we tested sympathy as well as sentiment. Combining all (crowdsourced labeled) data together, we tested whether sympathy labels were correlated with retweet count. We found a weak yet significant negative correlation (Spearman ($\rho$) =-0.06, p$<$0.05). Given the low correlation, we cannot state that such sympathetic tweets would result in higher information propagation. We further tested whether sentiment correlates with tweet spread, however we did not find a significant effect (Spearman ($\rho$) =0.011, p$=$5.78). While these findings may be in contrast to what Hanson et al. \cite{Hansen2011} and to what Berger et al. \cite{Berger2012} found, where neither sympathy nor sentiment increased virality, this could be due to hidden factors such as the timeliness and informativeness of the crisis news tweets in our dataset.

\begin{table}[t]
\centering
\scriptsize
{\begin{tabular}{lcc}
  \hline
 & Paris & Beirut \\ 
  \hline
Arab &  0.609  (N=708) & 0.976 (N=338)    \\ 
Western & 0.215 (N=4270) & 0.152$^{\dagger}$ (N=112)  \\ 

 \midrule\\[-1.5ex]

   \multicolumn{3}{c}{$^{\dagger}$Value drawn from labeled (ground truth) data.}  \\
   
\end{tabular}}

\caption{Ratio of sympathetic tweets.}
\label{table:ratio_positive}
\end{table}

What is interesting to observe here is that while retweeting behavior appears to be impartial as to whether a tweet is sympathetic or not, it does appear (within our dataset) that this similarly applies to sentiment labels, which is in contrast to Berger et al.'s findings that emotionality enhances virality. This is likely attributed to changing user information needs in the days following an attack (e.g., retweeting cries for help). 

\section{Discussion}

\subsection{Foreign News Coverage Differences}

Examining differences in foreign news coverage is important as the media has the power to shape the public perception about foreign countries \cite{Wanta2004}. Not just selection, but how a piece of news is reported has the capacity to evoke compassion, which could lead to various charitable acts, such as fund-raising to provide monetary support \cite{Kwak2014}. When an unexpected crisis such as a terror attack on Beirut or Paris strikes, with quality coverage, it has the power to instigate collective worldwide public action. The unique context of our study, marked by the rare successive occurrences of two crisis events at places within the Western and Arab world, allows exploring the relative differences specific to those two worlds. We believe looking into this closely is especially important in this era as tensions between those two worlds intensify in various regions and the need to deepen mutual understanding increases. 

As mentioned earlier, the issue of differential coverage was picked up by media and speculations were made that the Beirut attacks received less attention than Paris. Our analysis supports the speculation that at least on Twitter, indeed there were less coverage of Beirut (RQ1). It may be obvious that the Paris attacks were more newsworthy than the Beirut attacks; this is what news flow theory would predict \cite{Segev2015,Wu2000}, since Paris fulfills the criteria of a familiar, powerful foreign country, close geographically to other European states, and had an unexpected, human-induced crisis occur. This is also what the newsworthiness theory by Galtung and Holmboe \cite{Galtung1965} would predict, as newsworthiness depends on factors such as frequency, intensity, unambiguity, and unexpectedness. Although such theories offer explanations about coverage bias, one can draw different opinions about the news value considering the contemporary context of the two worlds (e.g., perhaps Arab states have heightened empathy towards other Arab states), and we believe our study provides one possible way of framing the discussion; how the media of two worlds allocated attention to the event abroad in relation to the domestic event. This framing can be further extended to coverage of other events and other countries, opening a space for future work. 

While our main finding was that the media of both worlds produced less amount of sympathetic coverage for a crisis abroad (RQ2), an additional observation was that Arab media coverage was more sympathetic, even though we expected that Western media would be more sympathetic when covering Paris. We interpret this result interesting as it could be capturing cultural differences in journalism practice \cite{deuze2002national,hanitzsch2011mapping}. We believe it is important to understand and consider cultural differences as it can enable more sophisticated analyses of biases in the media of different cultural regions. This result also suggests the possibility of approaching cultural differences in journalism practice using advanced computational methods. As we applied a deep learning technique to detect sympathetic tweets, we provide an initial step towards further large-scale quantitative analysis techniques that can shed light on capturing such cultural differences. As an example, while we did not observe differences in how sympathetic news tweets spread by Twitter users (RQ3), we found that the best predictor of retweeting activity is the number of followers (Spearman ($\rho$) = 0.76, p$<$0.001). While this fits our intuitions and understanding of how Twitter works \cite{Cha10}, it invites further thought on how network effects manifest across content and global events. 

Relatedly, as platforms such as Twitter are widely used for journalism at a global scale, our work also invites the development of tools or design features to support reflections on the newsworthiness of international events. For example, a dynamic dashboard to track foreign news topics covered in the media of different countries. From an end-user perspective, our work furthermore contributes a first step towards an approach that can be implemented as a Twitter feature that automatically detects not just sentiment, but also sympathy. By accounting for time-criticality in news reporting during crisis situations, this can can aid journalistic practice in situ, by automatically detecting sympathy in Twitter text as news reporters rush to formulate event-specific content. Finally, while we did not test this, it is relevant to investigate whether/how English and non-English language media differ in reporting.

\vspace{-4px}

\subsection{Limitations}

We have analyzed a snapshot of Twitter data after the Beirut and Paris attacks, where we made several decisions in how we treated the data. Also, it was not viable to test whether news sympathy differed across all countries in the Western world, and compare with news coverage of all Arab countries. Instead, we focused on geographic regions that made sense to test in the context of these attacks, and in this regard, our work provides an approximation of the differences in news coverage across Western and Arab media. Furthermore, our hashtags for the Beirut dataset were written in Latin alphabet. However, we were able to collect a sufficient number of tweets to run our analyses, where 62.1\% of Arab media tweets in our final Beirut dataset consisted of Arabic tweets.

A potential limiting factor is the `hostile media effect' \cite{Perloff2015}, a perceptual theory of mass communication that refers to the tendency for individuals with strong preexisting attitudes on an issue to perceive neutral media coverage as biased against their perspective, and instead adopt the antagonists' point of view. This phenomenon could have surfaced in crowdworkers, where workers from one country could have held biases when rating tweets. This may be amplified since we did not conceal names of news accounts (which we did to ensure tweets were close to real-world conditions). However, since each tweet received at least three trusted judgments, we have confidence that our annotations are likely to be untainted by such a bias. One way to address this is to survey crowdworkers for demographics and political orientation (cf., \cite{Mason2012}). Finally, we looked only at Twitter, which is shown to contain network biases \cite{Gonzalez2014}. Nevertheless, the large quantity of tweets and prevalence of reactions on this divided issue of Arab versus Western news coverage provided an initial step to test our approach.

\section{Conclusion and Future Work}

We presented a data-driven approach to tease out differences between Western and Arab Twitter news reporting of the 2015 Paris and Beirut attacks, where we found evidence for differential coverage across the attacks. For sympathy bias, we found that Western media tweets were less sympathetic when covering the Beirut attacks, however Western media was overall less sympathetic than Arab media, even for Paris. Finally, based on our labeled data, we trained a deep CNN to predict sympathy from unlabeled data, and results further supported our ground truth analysis that Western media had less sympathetic tweets than the Arab media, across both attacks. As a more general framework, our work contributes to an understanding of media bias on Twitter, and factors that may influence it, which are not necessarily limited to the studied attacks. We believe the methods we adopted are more widely applicable to other areas of computational journalism, and can serve as useful tools to better understand, expose, and design around media bias. 

\section{Acknowledgments}

This research was supported in part by the Volkswagen Foundation through a Lichtenberg Professorship.

%
%
%
%
%
\balance{}

\bibliographystyle{SIGCHI-Reference-Format}
\bibliography{crisisnewssympathy}

\end{document}